# Epitaxial growth of hexagonal tungsten bronze $Cs_xWO_3$ films in superconducting phase region exceeding bulk limit

Takuto Soma[1], Kohei Yoshimatsu[1, a)], and Akira Ohtomo[1,2]

[1]*Department of Chemical Science and Engineering, Tokyo Institute of Technology, 2-12-1 Ookayama, Meguro, Tokyo 152-8552, Japan*

[2]*Materials Research Center for Element Strategy (MCES), Tokyo Institute of Technology, Yokohama 226-8503, Japan*

a) electronic mail: k-yoshi@apc.titech.ac.jp

We report epitaxial synthesis of superconducting $Cs_xWO_3$ ($x$ = 0.11, 0.20, 0.31) films on Y-stabilized $ZrO_2$ (111) substrates. The hexagonal crystal structure was verified not only for composition within the stable region of bulk ($x$ = 0.20, 0.31), but also for the out-of-range composition ($x$ = 0.11). The onset of superconducting transition temperature ($T_C$) was recorded 5.8 K for $x$ = 0.11. We found strong correlation between $T_C$ and $c$-axis length, irrespective of the Cs content. The results indicate that hidden superconducting phase region of hexagonal tungsten bronze is accessible by using epitaxial synthesis of lightly doped films.

Tungsten bronzes $A_x$WO$_3$ have been synthesized with a wide variety of $A$-site ions (H$^+$, Li$^+$, Na$^+$, K$^+$, Rb$^+$, Cs$^+$, Ca$^{2+}$, In$^+$, and complex ions such as NH$_4^+$).[1,2] Depending on the $A$-site ions and their content $x$, they take various crystal structures having corner-shared frames of WO$_6$ octahedra. When large alkali metals such as K, Rb, or Cs are doped, $A_x$WO$_3$ with a hexagonal structure, *i.e.*, hexagonal tungsten bronze (HTB), is stably formed in an appropriate range of $x$.[3–5] The HTB has characteristic vacant sites along the $c$-axis surrounded by six WO$_6$ octahedra [Fig. 1(a)], which are regarded as channels for intercalation of smaller alkali metals into the hosts.[6] In fact, the HTB has been investigated for cathode materials of Li-ion batteries.

The HTBs have also attracted much attention in terms of their physical properties. The mixed valance in W ions, $A$-site nonstoichiometry, and characteristic structures provide an ideal platform to investigate exotic properties. In particular, the HTBs exhibit superconductivity and their superconducting transition temperature ($T_C$) tends to increase as $x$ decreases.[3–5,7,8] In bulk Cs$_x$WO$_3$, superconductivity completely vanishes when $x$ exceeds either lower or upper bounds of HTB-phase region; $T_C$ increases from 1.0 K ($x = 0.33$) to 6.7 K ($x = 0.19$),[4] which is the highest among HTBs. The upper bound at $x = 0.33$ corresponds to a crystallographic limit, where all the $A$ sites are filled, while the lower bound corresponds to a structural stability limit of the HTB phase ($x$ = 0.17 for K and 0.16 for Rb).[3,5] Therefore, optimal *A*-site doping in HTB has been thought to be hidden by the structural stability limit.[9]

The epitaxial synthesis of metastable compounds is a powerful approach to pave the way for exploring the exotic properties. It is worth testing this approach for reaching the hidden superconducting states in the HTBs. Wu *et al*. first fabricated superconducting hexagonal $K_{0.33}WO_3$ films on $LaAlO_3$ and Y-stabilized $ZrO_2$ (YSZ) substrates by using pulsed-laser deposition (PLD).[10] In recent, undoped hexagonal $WO_3$ films were epitaxially synthesized by using sputtering and MBE,[11] but superconductivity did not appear in the as-grown and ionic-liquid-gated samples.

In this study, we have fabricated the *c*-axis oriented hexagonal $Cs_xWO_3$ ($x$ = 0.11, 0.20, 0.31) epitaxial films on YSZ (111) substrates. All the films, even having thermodynamically unstable composition ($x$ = 0.11), exhibit superconductivity with $T_C$ comparable to bulk. We find that post-annealing is essential to improvement of $T_C$. Moreover, we show that the *c*-axis length is a crucial factor for the superconducting properties, rather than Cs content.

Several hundreds-nm-thick $Cs_xWO_3$ films were grown on YSZ (111) substrates by using PLD with a KrF excimer laser (0.6 J/cm$^2$). The ceramic tablets were prepared by conventional solid-state reaction steps, starting from mixing $Cs_2CO_3$ and $WO_3$ powders with different compositions (Cs/W = 0.10, 0.20, 0.33).  The substrate temperature was set 750°C and oxygen pressure ($P_{O2}$) was controlled in a range of 5−50 mTorr by continuous flow of pure oxygen (6 N purity).  After the growth, the films were cooled to RT with keeping $P_{O2}$.  Some of the films were annealed in $P_{O2}$ ranging from $10^{-8}$ Torr to 0.1 mTorr.  The annealing temperature and duration were varied in a range of 700−750°C and 2−34 h, respectively.

The film composition was analyzed by using a scanning electron microscopy equipped with an electron probe microanalyzer (EPMA).  The Cs contents in the as-grown and annealed films were $x$ = 0.11, 0.20, and 0.31, as each target was used. The crystal structures and epitaxial relationship were investigated by a laboratory X-ray diffraction (XRD) apparatus with Cu K$\alpha_1$ radiation.  Temperature dependence of resistivity was measured by a standard four-probe method using a physical property measurement system (Quantum Design, PPMS).

Figures 1(a) and 1(b) show schematic illustration of crystal structures of $Cs_xWO_3$ projected along the [001] direction and YSZ projected along the [111] direction, respectively.  The *a*-axis lattice constant (7.42 Å) of hexagonal $Cs_{0.2}WO_3$ is close to double $d_{1\bar{1}0}$ spacing (7.27 Å) of YSZ, and their mismatch is only 2.1%.[12,13)

Consequently, the $c$-axis oriented HTB films were obtained when a mixed valence state of $W^{5+}/W^{6+}$ was accommodated under appropriate $P_{O2}$.  Figure 2 (a) shows out-of-plane XRD patterns of $Cs_{0.20}WO_3$ films grown in various $P_{O2}$.  For all the films, HTB 002 and 004 reflections were detected at $2\theta \approx 23°$ and 48°, respectively.  The $c$-axis length for the film grown in $P_{O2}$ = 10 mTorr was estimated to be 7.56 Å, which was close to that of bulk ($c$ = 7.57 Å).  However, small reflections of secondary phases (marked with filled triangles) were also detected except for $P_{O2}$ = 10 mTorr.  The reflections at $2\theta \approx 31°$ and 46° seen for the film grown in $P_{O2}$ = 50 mTorr can be attributed to fully oxidized compounds such as $Cs_2W^{6+}O_4$ and $W^{6+}O_3$.  Correspondingly, the film was insulating.  In contrast, a sharp peak at $2\theta \approx 37°$ seen for the film grown in $P_{O2}$ = 5 mTorr was assigned to 020 reflection of rutile-type structure $W^{4+}O_2$.[14)  These results reflect subtle oxygen nonstoichiometry arising from similar redox potentials of tungsten-ion species.

Despite the thermodynamically unstable phase, HTB films with $x$ = 0.11 could be grown in $P_{O2}$ = 10 mTorr [Fig 2 (b)].  The $c$-axis length was estimated to be 7.55 Å, slightly smaller than that of the $Cs_{0.20}WO_3$ film.  We also measured asymmetric $Cs_xWO_3$ 202 reflection to evaluate the $a$-axis length as 7.44 Å for $x$ = 0.11 and 7.43 Å for $x = 0.20$. The latter was close to that of bulk ($a = 7.42$ Å),[13] indicating strain-free and relaxing HTB films.

We carried out XRD $\varphi$-scan for asymmetric reflections to confirm the epitaxial relationship just as shown in Fig. 1. The $Cs_xWO_3$ 202 and YSZ 100 reflections showed six- and three-fold symmetry, respectively [Fig. 2 (c)]. Thus, epitaxial relationship of $Cs_xWO_3$ [100] ∥ YSZ [1$\bar{1}$0] was verified.

Figure 3(a) shows temperature dependence of resistivity for $Cs_xWO_3$ films ($x = 0, 0.11, 0.20, 0.31$). The $WO_3$ ($x = 0$) film has a distorted $ReO_3$-type structure, details of which are described elsewhere.[15] Starting from fairly high resistivity of $x = 0$, temperature dependent resistivity systematically decreased with increasing $x$, exhibiting the insulator to metal transition (IMT) at $0.11 < x < 0.20$. We would like to emphasize that IMT has not been reported for hexagonal $Cs_xWO_3$ thus far (although IMT occurs in tetragonal and/or cubic $Li_xWO_3$ at $0.2 < x < 0.24$ and in cubic $Na_xWO_3$ at $0.23 < x < 0.29$[15,16]). Furthermore, superconductivity was observed at low temperatures for the films with $x = 0.11$ and 0.20. We also would like to emphasize that superconductivity in $Cs_{0.11}WO_3$ has been observed for the first time. In contrast, the superconducting transition was not observed for the $Cs_{0.31}WO_3$ film above 1.9 K, which was consistent with the previous reports.[4]

We sought other factors that influenced the superconducting properties. As shown in Figs. 3(b) and 3(c), the as-grown films with $x = 0.11$ and 0.20 exhibited the onset of $T_C$ ($T_{C, onset}$) at 5.5 and 4.9 K, respectively. We note that $T_{C, onset}$ for $x = 0.20$ is considerably lower than $T_C$ for bulk $Cs_{0.2}WO_3$ (6.4 K). We attributed this discrepancy to deviation from the oxygen stoichiometry. For $Rb_xWO_3$, strong oxygen-content dependence of $T_C$ was reported.[17] We conducted annealing in $P_{O2} \approx 0.1$ mTorr to tune the oxygen stoichiometry carefully. The post-annealed films exhibited sharp transitions with higher $T_{C, onset}$ (5.8 K and 5.4 K for $x = 0.11$ and 0.20, respectively). The crystallinity of the films (namely, FWHM of the XRD $\omega$-scans) hardly changed after the annealing. In contrast, the $c$-axis length somewhat varied in a systematic fashion as will be discussed later. These variations were due to neither the lack of Cs content nor stain relaxation as described already. Therefore, we presume that lattice parameter is a crucial factor for the superconducting properties.

We tested this hypothesis for the non-superconducting $Cs_{0.31}WO_3$ film. The sample was first annealed at 750°C in $P_{O2} \approx 0.1$ mTorr for extended durations. Surprisingly, superconductivity was observed; $T_{C, onset}$ increased from 3.5 K (after 4 h) to 4.6 K (after 16 h) and then was saturated at around 4.8 K (after 34 h). Meanwhile, the $c$-axis length drastically decreased and became comparable to that of bulk $Cs_{0.2}WO_3$ ($\approx$ 7.57 Å). On the other hand, the film showed no transition and the $c$-axis length close to that of bulk $Cs_{0.33}WO_3$ after reducing in Ar/H$_2$ mixed gas (1vol% H$_2$) at 750°C. Moreover, superconductivity ($T_{C, onset}$ = 3.0 K) was observed again after the annealing at 750°C in $P_{O2} \approx$ 0.1 mTorr. These results indicate that Cs doping is not a primary factor for modulation of the superconducting properties.

Now we propose a new superconducting phase diagram as shown in Fig. 4. $T_{C, onset}$ are plotted as a function of the $c$-axis length for our $Cs_xWO_3$ films. The conventional phase diagram of bulk $Cs_xWO_3$ is also shown.[4] As for bulk, $T_C$ is scalable to both Cs content $x$ and the $c$-axis length because $c$ vs. $x$ obeys Vegard's law.[13] Here the structural stability limit of bulk HTB phase can be located at $c$ = 7.567Å (indicated by broken line). Our $Cs_xWO_3$ films follow similar trend, irrespective of $x$, but those having smaller $c$-axis length ($x$ = 0.11, 0.20) exceed the bulk limit.

This unusual phase stability is likely brought by nonequilibrium nature of PLD as well as the epitaxial stabilization effect. Therefore, further optimization of PLD growth and annealing conditions may enhance $T_C$. Moreover, the present study provides further insight into research on tungsten bronzes, when combined with a number of approaches. For example, indium content in HTB $In_xWO_3$ can be reduced to $x$ = 0.11 without the structural phase transition by annealing with iodine.[9] The other techniques such as electrostatic doping[18] and electrochemical deintercalation[19] in liquid electrolytes will be usable.

In summary, we have grown hexagonal $Cs_xWO_3$ ($x$ = 0.11, 0.20, 0.31) epitaxial films on YSZ (111) substrates by using PLD. The transport and XRD measurements revealed that superconductivity emerged even out of structural stability region of HTB $Cs_xWO_3$ bulk. By annealing under various conditions, a number of samples were prepared to elucidate strong correlation between $T_C$ and the $c$-axis length. We concluded that superconducting properties of HTBs $Cs_xWO_3$ were governed by lattice parameters rather than Cs content. These results strongly suggest that the epitaxial stabilization is a powerful approach for studying the HTB system.

**Acknowledgments** The authors thank Y. Wada, M. Maitani, and S. Tsubaki for assistance regarding EPMA. This work was partly supported by MEXT Elements Strategy Initiative to Form Core Research Center and a Grant-in-Aid for Scientific Research (Nos. 15H03881 and 16H05983).

**References**

1) J. B. Goodenough, Prog. Solid State Chem. **5**, 145 (1971).

2) C. G. Granqvist, Sol. Energy Mater. Sol. Cells **60**, 201 (2000).

3) R. K. Stanley, R. C. Morris, and W. G. Moulton, Phys. Rev. B **20**, 1903 (1979).

4) M. R. Skokan, W. G. Moulton, and R. C. Morris, Phys. Rev. B **20**, 3670 (1979).

5) L. H. Cadwell, R. C. Morris, and W. G. Moulton, Phys. Rev. B **23**, 2219 (1981).

6) Q. Zhong and K. Colbow, Thin Solid Films **196**, 305 (1991).

7) P. E. Bierstedt, T. A. Bither, and F. J. Darnell, Solid State Commun. **4**, 25 (1966).

8) N. Haldolaarachchige, Q. Gibson, J. Krizan, and R. J. Cava, Phys. Rev. B **89**, 104520 (2014).

9) J. D. Bocarsly, D. Hirai, M. N. Ali, and R. J. Cava, Europhys. Lett. **103**, 17001 (2013).

10) P. M. Wu, C. Hart, K. Luna, K. Munakata, A. Tsukada, S. H. Risbud, T. H. Geballe, and M. R. Beasley, Phys. Rev. B **89**, 184501 (2014).

11) P. M. Wu, S. Ishii, K. Tanabe, K. Munakata, R. H. Hammond, K. Tokiwa, T. H. Geballe, and M. R. Beasley, Appl. Phys. Lett. **106**, 042602 (2015).

12) T. H. Etsell and S. N. Flengas, Chem. Rev. **70,** 339 (1970).

13) A. Hussain, Acta Chemica Scandinavica A **32**, 479 (1978).

14) F. J. Wong and S. Ramanathan, J. Mater. Res. **28** 2555 (2013).

15) K. Yoshimatsu, T. Soma, and A. Ohtomo, submitted to Appl. Phys. Lett.

16) P. A. Lightsey, D. A. Lilienfeld, and D. F. Holcomb, Phys. Rev. B **14**, 4730 (1976).

17) D. C. Ling, Y. C. Shao, J. W. Chiou, W. F. Pong, S. H. Wu, Y. Y. Chen, and F. Z. Chien, J. Phys.: Conf. Ser. **150**, 052141 (2009).

18) J. T. Ye, Y. J. Zhang, R. Akashi, M. S. Bahramy, R. Arita, and Y. Iwasa, Science **338**, 1193 (2012).

19) K. Yoshimatsu, M. Niwa, H. Mashiko, T. Oshima, and A. Ohtomo, Sci. Rep. **5**, 16325 (2015).

**Figure Captions**

**Fig. 1.** Schematic illustration of crystal structures of (a) Cs$_x$WO$_3$ projected along [001] and (b) YSZ projected along [111] directions. The red and green spheres indicate oxygen and Cs atoms, respectively. The hatched areas surrounded by broken lines represent the surface unit cells for epitaxial growth.

**Fig. 2.** XRD patterns of Cs$_x$WO$_3$ films and YSZ substrates. (a) Out-of-plane XRD patterns for Cs$_{0.20}$WO$_3$ films grown in various $P_{O2}$. (b) Out-of-plane XRD pattern of Cs$_{0.11}$WO$_3$ films grown in $P_{O2}$ = 10 mTorr. (c) XRD $\varphi$-scans of Cs$_x$WO$_3$ 202 and YSZ 111 reflections. Filled and open triangles indicate reflections coming from secondary phases and sample stage, respectively.

**Fig. 3.** (a) Temperature dependence of resistivity for Cs$_x$WO$_3$ films ($x$ = 0.11, 0.20, 0.31) and a pure WO$_3$ film. Temperature dependence of normalized resistivity for as-grown and annealed films of (b) Cs$_{0.20}$WO$_3$ and (c) Cs$_{0.11}$WO$_3$.

**Fig. 4.** Superconducting transition temperatures as a function of the $c$-axis length for as-grown and annealed Cs$_x$WO$_3$ films. The data for bulk are also plotted for comparison. The bulk references are rearranged from the $x$ dependencies of $T_\mathrm{C}$ and $c$, reported independently in Ref. 4 and 13, assuming Vegard's law for $c$ vs. $x$. The broken lines indicate guide for eyes.

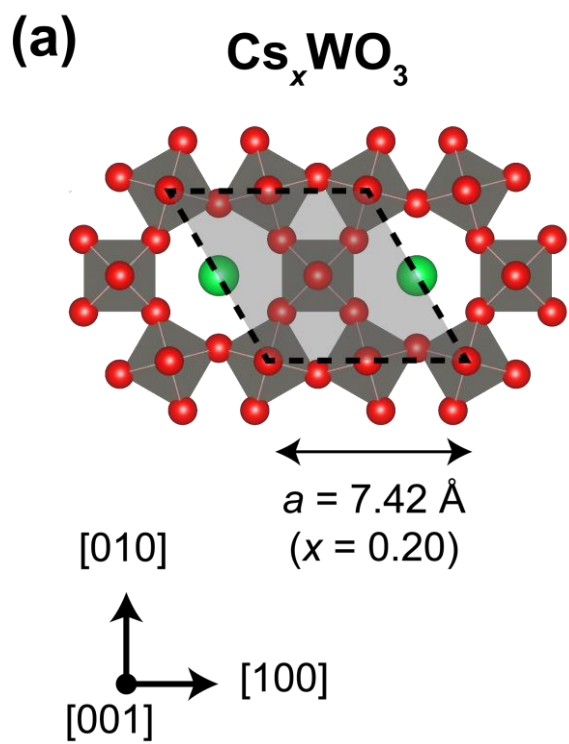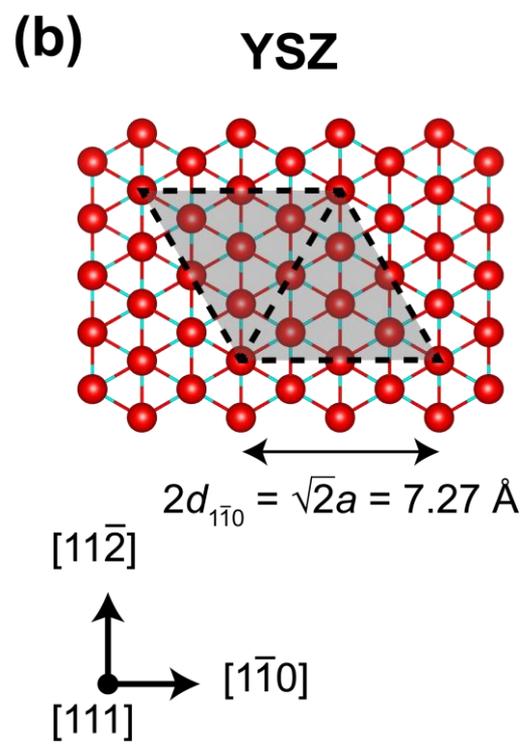

Figure 1 T. Soma *et al*.

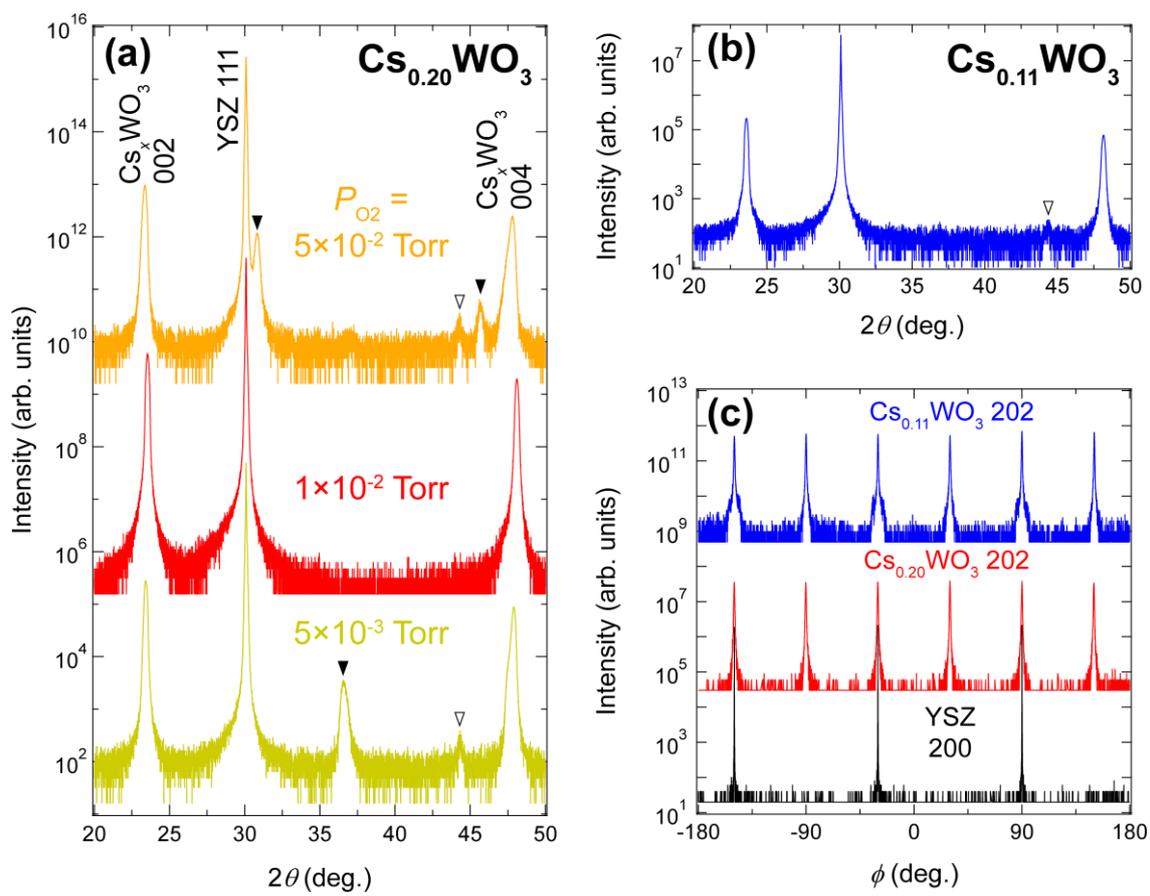

Figure 2 T. Soma *et al*.

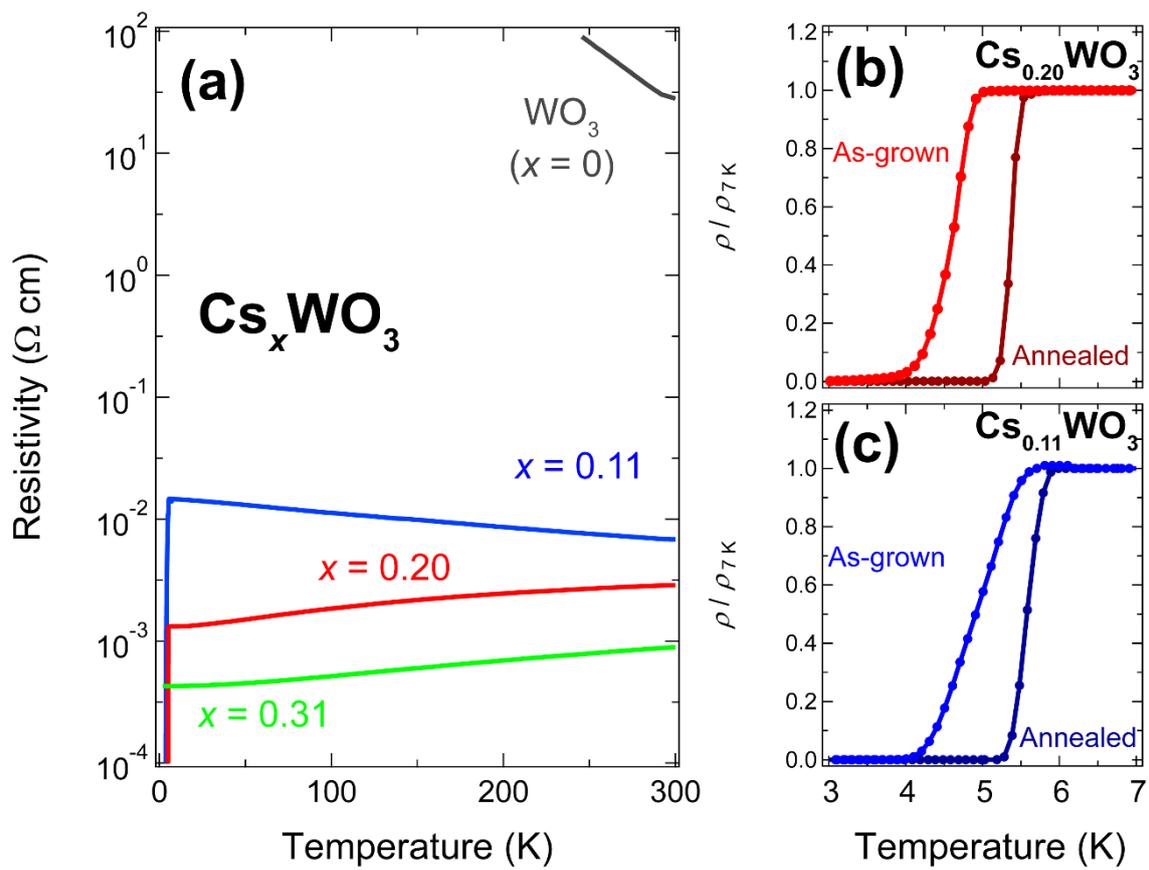

Figure 3 T. Soma *et al*.

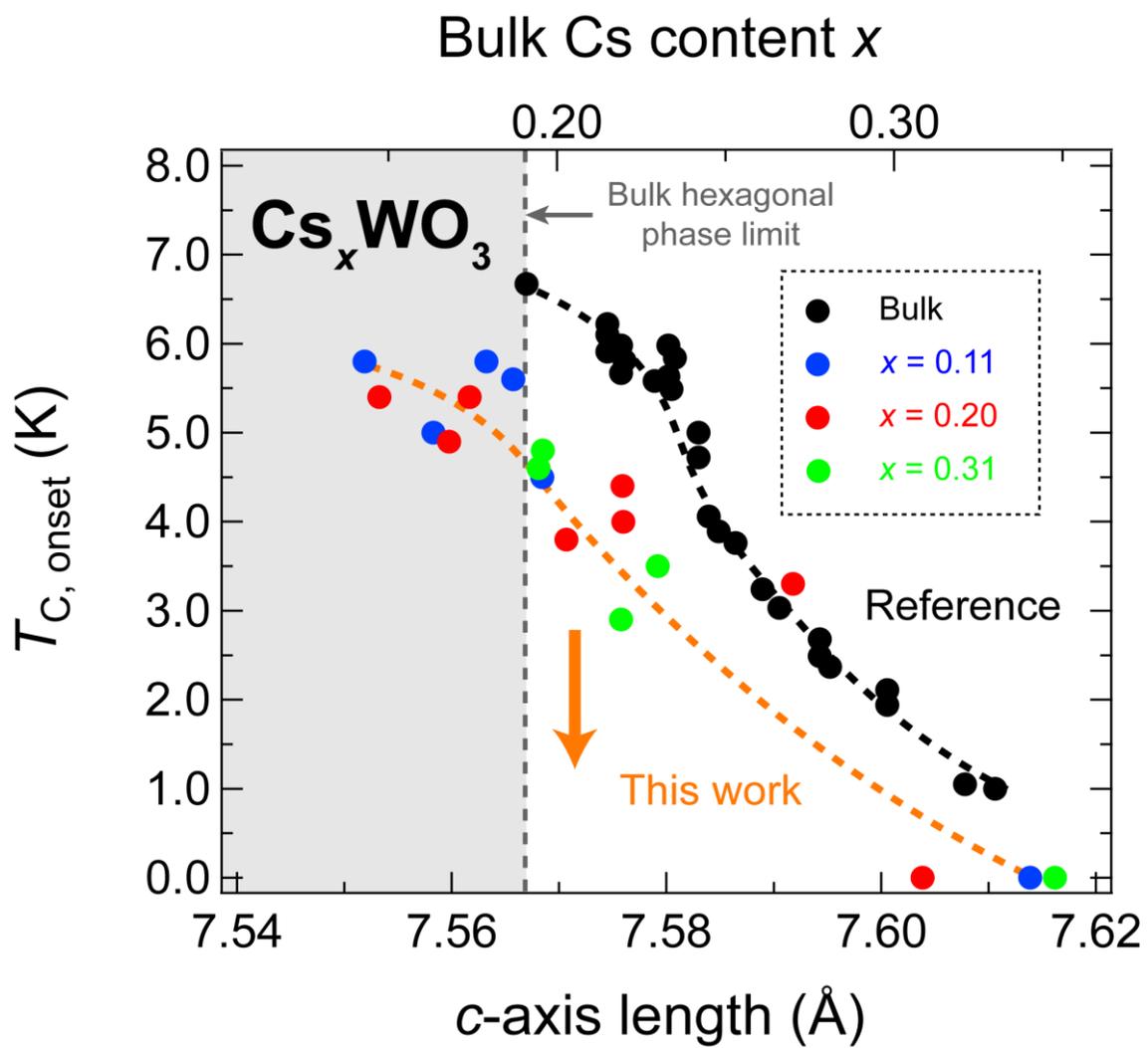

Figure 4 T. Soma *et al*.